\documentclass[12pt]{article}

\usepackage{amssymb,amsbsy,amsthm,latexsym,amsmath}

\newcommand{\dd}{{\rm d}}
\newcommand{\ee}{{\rm e}}
\newcommand{\ii}{{\rm i}}

\theoremstyle{plain}

\newtheorem{thm}{Theorem}[section]
\newtheorem{lemma}[thm]{Lemma}
\newtheorem{cor}[thm]{Corollary}
\newtheorem{definition}[thm]{Definition}
\newtheorem{prop}[thm]{Proposition}
\newtheorem{rmk}[thm]{Remark}

\newcounter{saveeqn}
\newcounter{App} 
\newcommand{\app}{%
\stepcounter{App}%
\setcounter{saveeqn}{\value{equation}}
\setcounter{equation}{0}%
\renewcommand{\theequation}{\Alph{App}\arabic{equation}} 
\renewcommand{\thesection}{\Alph{App}} }
\newcommand{\appende}{%
\setcounter{equation}{\value{saveeqn}}%
\renewcommand{\theequation}{\arabic{equation}}  }

\begin{document}

\begin{flushright}
\vspace{.4cm}
\today
\end{flushright}

\vspace{.4cm}

\begin{center}

\noindent{\Large\bf Generalized local interactions in 1D: 
solutions of quantum many-body systems describing distinguishable
particles}

\vspace{1 cm}

\noindent{\large Martin Halln\"as, Edwin Langmann and Cornelius
Paufler}\\

\vspace{0.3 cm}

\noindent{\em Mathematical Physics, Department of Physics, KTH-SCFAB, SE-106 91
Stockholm, Sweden}\\
\end{center}

\vspace{0.3 cm}

\begin{abstract}
As is well-known, there exists a four parameter family of local
interactions in 1D.  We interpret these parameters as coupling
constants of delta-type interactions which include different kinds of
momentum dependent terms, and we determine all cases leading to
many-body systems of {\em distinguishable} particles which are exactly
solvable by the coordinate Bethe Ansatz.  We find two such families of
systems, one with two independent coupling constants deforming the
well-known delta interaction model to non-identical particles, and the
other with a particular one-parameter combination of the delta- and
(so-called) delta-prime interaction.  We also find that the model of
non-identical particles gives rise to a somewhat unusual solution of
the Yang-Baxter relations.  For the other model we write down explicit
formulas for all eigenfunctions.
\end{abstract}

\section{Introduction}
By general physical arguments one expects that, to understand the
low-energy properties of a quantum system, one can ignore details of
the short distance structure of the interactions and replace them by
local interactions which are singular and non-trivial only at a
point. The most prominent such local interaction is formally defined
by a delta function potential and parameterized by {\em one} real
parameter, but it is known since quite some time that, in one
dimension (1D), the most general local interaction is characterized by
{\em four} real parameters \cite{S2,AGHH}.  While the earliest known
examples of exactly solvable 1D quantum many-body systems with
two-body interactions are defined with the delta function potential
\cite{LL,McG,Y} (see also \cite{Gaudin}), the full four parameter
family of local interactions has received much less attention in this
context until recently \cite{G,CNL,CS1,AFK,AFK1,GLP,CMR,HL} (see also
Chapter~7 in \cite{AK}).

In this paper we consider the 1D quantum many-body systems with
two-body interactions given by the general four parameter family of
local interactions, and we determine {\em all} cases which are exactly
solvable by the coordinate Bethe Ansatz even in the general case of
distinguishable particles. (The same problem was also studied in
\cite{AFK}, but our approach and results are different, as discussed
in more detail below.) We find and solve two families of such models
which provide interesting generalizations of previously known
cases. We use the parameterization of the general local interaction
proposed in \cite{EG}, and we suggest a natural physical
interpretation of these parameters as coupling constants of delta-type
interactions which also include different kinds of momentum dependent
terms. This allows us to write down formal Hamiltonians to define and
interpret these models in a simple manner.  While this interpretation
is different from others in the literature \cite{S2,CS3}, the
mathematically precise formulation of our models is the usual one in
terms of boundary conditions.  In the following two paragraphs we
describe in more detail the exactly solvable models which we find.

The first family of models can be formally defined by the
following Hamiltonian,
\begin{eqnarray}
\label{H1}
  H^{(1)} & = & -\sum_{j=1}^N\partial_{x_j}^2 + \sum_{j<k}{\Big\lbrack}
  2c\delta(x_j - x_k) + 2\eta \ii\left(\partial_{x_j} -
  \partial_{x_k}\right)\delta(x_j - x_k)\nonumber\\&&+\ 2\eta \delta(x_j - x_k)\
  \ii\left(\partial_{x_j} - \partial_{x_k}\right){\Big\rbrack}
\end{eqnarray}
which depends on two real coupling constants $c$ and $\eta$.  For this
and all other many-body models considered in this paper we assume that
the particles move on the full real line, $x_j\in\mathbb{R}$ for
$j=1,2,\ldots,N$, and we ignore bound states, which makes our solution
complete only under certain restrictions on the coupling constants,
such as $c>0$.\footnote{Bound states are of course interesting but
beyond the scope of this work.}  This model is a one-parameter
extension of the famous 1D delta gas solved, in the boson case, by
Lieb and Liniger \cite{LL}, and in the general case of distinguishable
particles by Yang using the coordinate Bethe Ansatz \cite{Y}.  We
interpret the extension as a particular momentum dependent
interaction.\footnote{Since $\hat p_j \equiv -\ii\partial_{x_j}$ are
the particle momentum operators.} We find that it is possible to
generalize Yang's solution to the full model, even though it describes
identical particles only in the Yang case $\eta=0$ (since the
interaction terms with coupling $\eta$ are not invariant under
particle exchanges). 

A formal definition of the second model is
\begin{equation}
\label{H2}
  H^{(2)} = -\sum_{j=1}^N\partial_{x_j}^2 + \sum_{j<k}{\Big\lbrack}
  2c\delta(x_j - x_k) + \frac{2}{c}\left(\partial_{x_j} -
  \partial_{x_k}\right)\delta(x_j - x_k)\left(\partial_{x_j} -
  \partial_{x_k}\right){\Big\rbrack}
\end{equation}
with only one real coupling parameter $c$. The first interaction term
is identical to that of the 1D delta gas, but there is an additional
momentum dependent interaction term with inverse coupling strength. As
will be discussed below, this second interaction is identical with
what is usually referred to as delta-prime interaction
\cite{AGHH,S1}. This model describes identical particles, and it is
thus possible to restrict it to bosons and fermions. It is interesting
to note that, for fermions, the delta interaction is invisible (Pauli
principle), and the model reduces to one first solved by Cheon and
Shigehara \cite{CS1}, whereas for bosons the second interaction is
invisible reducing it to the 1D Bose gas solved in \cite{LL} (see also
\cite{GLP}). We find that the explicit eigenfunctions of this model
can be constructed in a remarkably simple manner by taking the
well-known eigenfunctions of the boson delta gas restricted to the
fundamental wedge $x_1<x_2<\ldots <x_N$ and extending it to all other
wedges $x_{Q1}<x_{Q2}<\ldots <x_{QN}$, $Q\in S_N$, using whatever
particle statistics one considers.  This generalizes the duality found
in \cite{CS1} to arbitrary exchange statistics.

We now recall some basic facts about general local interactions in 1D
(see e.g.\ \cite{EG} for a more comprehensive discussion).  The
simplest example of a model with such interactions can be formally
defined by a Hamiltonian
\begin{equation}
  H = -\partial_x^2 + \hat V
\end{equation}
with $x\in\mathbb{R}$ the particle coordinate and $\hat V$ an
interaction localized at $x = 0$, i.e., the action on wave functions
$\psi$ vanishes except at the origin. This implies that the
Schr\"odinger equation determining the eigenstates $\psi$ of $H$ is
trivial nearly everywhere, $\psi^{\prime\prime} + E\psi = 0$ for
$x\neq 0$, but the interaction results in non-trivial boundary
conditions at the singular point $x = 0$. Models of this type have
been studied extensively using the theory of defect indices, from
which it is known that the most general such interaction can be
parameterized by four real parameters \cite{S2,AGHH}. The most
prominent example is the delta function potential, 
\begin{equation}
\label{deltaInt}
  \hat V = c \delta(x), 
\end{equation}
parameterized by one real coupling constant $c$.  Another well known
special case is what often has been referred to as delta-prime
interaction; see Section I.4 of \cite{AGHH}.  Recently it was shown
that the boundary conditions defining this interaction arise from the
following momentum dependent potential,
\begin{equation}
\label{pdeltapInt}
  \hat V = 4\lambda \partial_x\delta(x)\partial_x
\end{equation}
with $\lambda$ a real coupling constant \cite{GLP}.\footnote{This
physical interpretation of the delta-prime interaction is similar to
the one of \v Seba \cite{S1} but it does not require any
renormalization of the coupling constant.} In this paper we extend
this point of view to the full four parameter family of local
interactions, and we show that they arise from the following momentum
dependent interaction,
\begin{equation}
\label{genInt}
  \hat V = c\delta(x) + 4\lambda\partial_x\delta(x)\partial_x +
  2(\gamma + \ii\eta)\partial_x\delta(x) - 2(\gamma -
  \ii\eta)\delta(x)\partial_x,
\end{equation}
where $c$, $\lambda$, $\gamma$ and $\eta$ are real coupling
parameters. This interpretation is different from the standard one as
a limit of conventional potentials which requires renormalizations of
the coupling constants \cite{CS3}. However, it leads to a convenient
parameterization $(c,\lambda,\gamma,\eta)$ of the local interactions
which has a natural physical interpretation and which is without
constraints.  Moreover, as shown in Section 2, by simple formal
computations the interaction in (\ref{genInt}) is turned into standard
boundary conditions which provide a mathematically rigorous
formulation of the model.

We can now give a more specific description of the many-body systems
we consider. They are defined by the Hamiltonian
\begin{equation}
  H_N = -\sum_{j=1}^N\partial_{x_j}^2 + \sum_{j<k}\hat
  V_{jk}\label{HN}
\end{equation}
with $x_j\in\mathbb{R}$ the particle coordinates and $\hat V_{jk}$
local two-body interactions obtained from the momentum dependent
potential in (\ref{genInt}) by replacing $x$ by the inter-particle
distance $x_j-x_k$. In our discussion of special cases of this model
we will use the notation $(c,\lambda,\gamma,\eta)$ introduced above,
e.g., $(c,0,0,\eta)$ refers to the model defined in (\ref{H1}).  It is
important to note that the general model $(c,\lambda,\gamma,\eta)$
describes identical particles only for $\eta=\gamma=0$ (since the
interaction terms $\partial_x\delta(x)\pm \delta(x)\partial_x$ in
(\ref{genInt}) change sign under particle exchanges $x\equiv
x_j-x_k\to -x$).  Thus, besides the delta-interaction model
$(c,0,0,0)$, also the model $(0,\lambda,0,0)$ is of special
interest. However, different from the former, the latter is {\em not}
exactly solvable in the general case of distinguishable particles
\cite{GLP}. Still, the restriction of this model to fermions is
interesting since it provides the only non-trivial exactly solvable
fermion model with local interactions in 1D (recall the the delta
interaction is trivial for fermions; note that we only discuss models
{\em without} internal degrees of freedom here). Moreover, it has a
natural physical interpretation as the non-relativistic limit of the
massive Thirring model \cite{GLP}. Our results in this paper imply
that the model in (\ref{H2}) is the only generalization of this
fermion model to distinguishable particles which remains exactly
solvable.

As already mentioned, the very same questions studied in this paper
was also studied by Albeverio {\em et.\ al.} \cite{AFK},\footnote{We
learnt about this work after finishing ours.} but their results differ
from ours. The reason for this is that in \cite{AFK} it is assumed
that the model describes identical particles so that the approach in
\cite{Y} applies, and this restricts their analysis (implicitly) to
the two parameter family $(c,\lambda,0,0)$ of local
interactions. Thus, while \cite{AFK} also finds that the model in
(\ref{H2}) is integrable, it concludes that the model (\ref{H1}) is
integrable only for the previously known case $\eta=0$.\footnote{In
\cite{AFK} a system is referred to as integrable if it is exactly
solvable by the coordinate Bethe Ansatz.}  Indeed, one can check that
the boundary conditions found to be integrable in \cite{AFK}, Eqs.\
(15) and (16), correspond to the cases $(c,0,0,0)$ and $(q,1/q,0,0)$
in our terminology.\footnote{The third case obtained in \cite{AFK}
corresponds to the limiting case $\lim_{\eta\to\infty}
(c\eta^2,0,\eta,0)$ in our terminology and thus violates the
assumptions under which it was obtained.}  Thus, our main result in
addition to \cite{AFK} is the extension of Yang's approach \cite{Y} to
models of non-identical particles and thereby giving the first
conclusive answer to the question posed in the second paragraph above.
Moreover, we also give a novel physical interpretation of these
models.

The remainder of the paper is organized as follows. In Section~2 we
discuss local interactions in 1D, and in particular the relation of
our interpretation to the standard one. In Section~3 we derive the
consistency conditions on the coordinate Bethe Ansatz for the
many-body systems with local interactions. Solving them we obtain the
two families of systems defined in (\ref{H1}) and (\ref{H2})
(Section~3.2). Section~3.3 contains the explicit solution of the
latter model. In Section~3.4 we construct the recursion relations for
the coefficients arising in the coordinate Bethe Ansatz for the model
in (\ref{H1}), which leads us to a somewhat unusual representation of
the Yang-Baxter relations.  We conclude with a few remarks on
interesting related models not considered in this paper
(Section~4). Some technical results related to our solution of the
Yang-Baxter relation are deferred to two appendices.

\section{Local interactions in 1D} Interactions localized at points in
1D have been studied extensively using the mathematical theory of
defect indices; see \cite{AGHH} and references therein. From these
studies it is well-known that the delta interaction is only one of
many possible local interactions, and that a general such interaction
can be characterized by four real coupling parameters. This can be
formally understood as follows: for a 1D Hamiltonian
$H=-\partial_x^2+\hat V$ with an interaction $\hat V$ localized at
$x=0$, all eigenfunctions $\psi(x)$ should be smooth everywhere except
at $x=0$, and $(H\psi)(x) = -\psi''(x)$ for non-zero $x$. Requiring
$H$ to be self-adjoint leads to the following consistency condition,
\begin{equation} 
	\int_{|x|>0}\dd x\, \Big( \overline{\phi''(x)} \psi(x)
	-\overline{\phi(x)} \psi''(x) \Big) = 0
\end{equation}
for arbitrary wave functions $\phi$ and $\psi$, or equivalently
\begin{equation} 
\label{cond}
	[\overline{\phi'}\psi - \overline{\phi} \psi']_{x=0^+} =
	[\overline{\phi'}\psi - \overline{\phi} \psi']_{x=-0^+} .
\end{equation}
General boundary conditions are of the form
\begin{eqnarray}
\label{bc} 
	\psi'(0^+) &=& u_{11} \psi'(-0^+) + u_{12}
	\psi(-0^+)\nonumber\\ \psi(0^+) &=& u_{21} \psi'(-0^+) +
	u_{22} \psi(-0^+)
\end{eqnarray}
(and similarly for $\phi$, of course) and are thus parameterized by
four complex parameters $u_{jk}$ which, upon imposing (\ref{cond}), are
reduced to two complex, or equivalently, four real parameters. 

One prominent example of such consistent boundary conditions are
\begin{eqnarray}
\label{bc1} 
	\psi(0^+) = \psi(-0^+)\nonumber\\ \psi'(0^+)-\psi'(-0^+) = c
	\psi(0^+) ,
\end{eqnarray}
which, as is well-known, corresponds to the delta interaction in
(\ref{deltaInt}). Another example is what is usually referred to as
delta-prime interaction, 
\begin{eqnarray}
\label{bc2} 
	\psi'(0^+) = \psi'(-0^+)\nonumber\\ \psi(0^+) - \psi(-0^+) =
	4\lambda \psi'(0^+) ,
\end{eqnarray}
and which corresponds to the momentum dependent interaction in
(\ref{pdeltapInt}).

As mentioned in the introduction, we use the following physical
interpretation of the full four parameter family of local
interactions,
\begin{equation} 
\label{H4} 
	H = -\partial_x^2 + \hat V,\quad \hat V = c \delta(x) +
	4\lambda\partial_x \delta(x)\partial_x + 2(\gamma +
	\ii\eta) \partial_x \delta(x) - 2(\gamma -
	\ii\eta) \delta(x) \partial_x,
\end{equation}
where $c$, $\lambda$, $\gamma$ and $\eta$ are real coupling
constants. Once realized, this result is rather plausible: the
operator $\hat V$ is obviously local and (formally) self-adjoint, and
it contains four real parameters. Moreover, it is obviously the most
general such interaction containing at most two derivatives, and it is
plausible that higher derivatives than that cannot lead to consistent
boundary conditions.  It is also not so difficult to show that it
indeed corresponds to the four parameter family of local interactions
introduced in \cite{S2,AGHH}: formally, the eigenvalue equation of $H$
in (\ref{H4}) is equal to
\begin{eqnarray} 
\label{H4psi} 
	-\psi''(x) + c \delta(x) \psi(0) + 4\lambda
	\delta^{\prime}(x)\psi^{\prime}(0) + 4\lambda
	\delta(x)\psi^{\prime\prime}(0) +\nonumber\\ + 2(\gamma +
	\ii\eta)\delta'(x)\psi(0) + 2(\gamma +
	\ii\eta)\delta(x)\psi^{\prime}(0) - 2(\gamma -
	\ii\eta)\delta(x) \psi'(0) = E\psi(x)
\end{eqnarray}
with $\psi$ the eigenfunction, $E$ the corresponding eigenvalue, and
primes indicating derivatives. The crucial point is that, due to the
singular interaction, the eigenfunctions are in general not continuous
at the singular point $x=0$, and one therefore has to interpret the
eigenfunction and its derivative at the singular point as the average
of the left- and right limits,
\begin{eqnarray} 
\label{psi0} 
	\psi(0) &\equiv & \frac12[\psi(0^+) + \psi(-0^+) ]\nonumber\\
        \psi'(0) &\equiv & \frac12[\psi'(0^+) + \psi'(-0^+)] . 
\end{eqnarray}
Using that (\ref{H4psi}) can be turned into boundary conditions of the
form (\ref{bc}) in the following way: integrating once from $x=-0^+$
to $x=0^+$, and twice, first from $x=-0^+$ to $x>0$ and then once more
from $x=-0^+$ to $x=0^+$ gives
\begin{eqnarray} 
\label{bcgeneral} 
	\psi'(0^+) -\psi'(-0^+) &=& c\psi(0) -2(\gamma -
	\ii\eta)\psi'(0) \nonumber\\ \psi(0^+) -\psi(-0^+) &=&
	4\lambda\psi'(0) + 2(\gamma + \ii\eta)\psi(0).
\end{eqnarray}
It is interesting to note that this parameterization of boundary
conditions is identical with the one proposed in \cite{EG}, Eq.\
(2.1).  Using (\ref{psi0}) we can write these latter equations as in
(\ref{bc}), with
\begin{eqnarray} 
\label{U} 
	U\equiv \left( \begin{matrix} u_{11}&u_{12}\\ u_{21}& u_{22}
	\end{matrix}\right) = (U_+)^{-1} U_- ,\quad U_\pm = \left(
	\begin{matrix} 1 \pm (\gamma-\ii\eta) & \mp c/2 \\ \mp 2\lambda &
	1\mp (\gamma+\ii\eta) \end{matrix}\right), 
\end{eqnarray} 
where we introduce a convenient matrix notation.\footnote{The special
cases $\det(U_+)=0$ where $U$ above is undefined are discussed in
\cite{EG}. We only mention that the case $\eta=\gamma=0$ and
$\lambda=1/c$ corresponds to separated boundary conditions, $\psi'(\pm
0^+) = c\psi(\pm 0)/2$.}  It is easily seen that the condition in
(\ref{cond}) is equivalent to
\begin{equation} 
\label{cond1} 	
	J \equiv \mbox{$\left( \begin{matrix} 0&-1\\ 1& 0
	\end{matrix}\right)$} = U^\dag J U ,
\end{equation} 
which indeed is satisfied by the matrix $U$ in (\ref{U}). We note that
this physical interpretation provides a convenient parameterization of
the full four parameter family of local interactions. In particular,
it is without constraints, in this respect distinguishing it from others used in
the
literature, e.g. \cite{S2,ADK}
\begin{eqnarray}
  U = \ee^{\ii\chi} \left( \begin{matrix} s & u \\ v & t
  \end{matrix}\right), \quad \chi\in[0,\pi), \quad
  s,t,u,v\in\mathbb{R} \; \mbox{ with } \; st-uv=1 .
\end{eqnarray} 
Rephrasing our result in this latter parameterization is
straightforward \cite{EG}, but its physical interpretation would be
less clear.

\section{Coordinate Bethe Ansatz}
In this section we consider the model of $N$ particles interacting via
the full four parameter family of local two-body interactions. It is
defined by the Hamiltonian $H_N$ in (\ref{HN}) with
\begin{eqnarray}
  \hat V_{jk} & = & 2c\delta(x_j - x_k) + 2\lambda\left(\partial_{x_j}
  - \partial_{x_k}\right)\delta(x_j - x_k)\left(\partial_{x_j} -
  \partial_{x_k}\right)\nonumber\\&&+\ 2(\gamma +
  \ii\eta)\left(\partial_{x_j} - \partial_{x_k}\right)
  - 2(\gamma - \ii\eta)\delta(x_j -
  x_k)\left(\partial_{x_j} - \partial_{x_k}\right),\label{Vjk}
\end{eqnarray}
which is the obvious $N$ particle generalization of
(\ref{genInt}). Using $2\partial_{x_j - x_k} = \partial_{x_j} -
\partial_{x_k}$ it is straightforward to generalize the boundary
conditions of the previous section to the $N$-particle case. We thus
deduce that the eigenfunctions $\psi$ of $H_N$ are defined by the free
Schr\"odinger equation
\begin{equation}
  \left(\sum_{j=1}^N\partial_j^2 +
  E\right)\psi\left(x_1,\ldots,x_N\right) = 0\quad\textrm{for}\quad
  x_j\neq x_k,\label{Scheq}
\end{equation}
and the following boundary conditions,
\begin{eqnarray}
  \label{boundCond}
  \left(\partial_{x_j} -
  \partial_{x_k}\right)\left\lbrack\psi\arrowvert_{x_j = x_k + 0^+} -
  \psi\arrowvert_{x_j = x_k - 0^+}\right\rbrack =
  2c\psi\arrowvert_{x_j = x_k} \\ \nonumber  -  
  2(\gamma-\ii\eta)\left(\partial_{x_j} -
  \partial_{x_k}\right)\psi\arrowvert_{x_j =
  x_k}\nonumber\\\psi\arrowvert_{x_j = x_k + 0^+} -
  \psi\arrowvert_{x_j = x_k - 0^+} = 2\lambda\left(\partial_{x_j} -
  \partial_{x_k}\right)\psi\arrowvert_{x_j = x_k} +
  2(\gamma+\ii\eta)\psi\arrowvert_{x_j = x_k},
\end{eqnarray}
where $j<k$ and where, similarly to the one particle case in the
previous section, the eigenfunctions as well as their derivatives must
be regularized at points of interaction $x_j = x_k$ as follows
\begin{eqnarray}
\label{regu}
	\psi\arrowvert_{x_j = x_k}\equiv \frac{1}{2}\left\lbrack
	\psi\arrowvert_{x_j = x_k + 0^+} + \psi\arrowvert_{x_j = x_k -
	0^+}\right\rbrack\nonumber\\ \left(\partial_{x_j} -
	\partial_{x_k}\right)\psi\arrowvert_{x_j = x_k}\equiv
	\frac{1}{2}\left(\partial_{x_j} -
	\partial_{x_k}\right)\left\lbrack \psi\arrowvert_{x_j = x_k +
	0^+} + \psi\arrowvert_{x_j = x_k - 0^+}\right\rbrack.
\end{eqnarray}
Equations (\ref{Scheq})--(\ref{regu}) provide a mathematically
rigorous formulation of our general model.

\subsection{Two particle case}
Before continuing our discussion of the full $N$ particle model we
consider in some detail the two particle case, $N = 2$, elucidating
some of the properties that make this model rather special.

The most important such property is that the Hamiltonian $H_2$ is
invariant under permutations of the coordinates $x_1$ and $x_2$ only
in the case $\gamma = \eta = 0$, and that we therefore, in general,
have a model of non-identical particles. To determine the implications
this has on the eigenfunctions we start by considering the following
Ansatz for the two particle scattering states,
\begin{equation}
\label{eig}
	\phi = \left\{\begin{array}{ll} \ee^{\ii k_1x_1 + \ii k_2x_2}
		+ S_R^+(k_1 - k_2)\ee^{\ii k_2x_1 + \ii k_1x_2}, &
		x_1<x_2\\ S_T^+(k_1 - k_2)\ee^{\ii k_1x_1 + \ii
		k_2x_2}, & x_2<x_1\end{array}\right.
\end{equation}
which, upon substitution into boundary conditions (\ref{boundCond}),
result in the following expressions for the two scattering amplitudes
$S_T^+$ and $S_R^+$,
\begin{eqnarray}
\label{scattAmpl}
	S_T^+(u) = \frac{(\gamma^2 + \eta^2 - 2\ii \eta + c\lambda -
	1)u}{\ii\lambda u^2 - (\gamma^2 + \eta^2 + c\lambda + 1)u -
	\ii c}\nonumber\\ S_R^+(u) = \frac{\ii\lambda u^2 + 2\gamma u +
	\ii c}{\ii\lambda u^2 - (\gamma^2 + \eta^2 + c\lambda + 1)u - \ii c},
\end{eqnarray}
where we have introduced $u = k_1 - k_2$. For future convenience we
introduce also the scattering amplitudes $S_R^-$ and $S_T^-$ obtained
from $S_R^+$ and $S_T^+$ by reverting the sign of $\gamma$ and
$\eta$,
\begin{equation}
\label{scattAmplMinus}
  S_R^- = S_R^+\arrowvert_{\gamma\rightarrow -\gamma,\eta\rightarrow
  -\eta},\quad S_T^- = S_T^+\arrowvert_{\gamma\rightarrow
  -\gamma,\eta\rightarrow -\eta}.
\end{equation}
{}From this Ansatz we construct a second set of eigenfunctions of
$H_2$ by noting that the Hamiltonian $H_2$, while not invariant under
the exchange of the coordinates $x_1$ and $x_2$, is invariant under
the simultaneous exchange of the coordinates $x_1$ and $x_2$ and
replacement of $\gamma$ and $\eta$ by $-\gamma$ and $-\eta$
respectively. This implies that also
$\phi\arrowvert_{x_1\leftrightarrow x_2, \gamma\rightarrow
-\gamma,\eta\rightarrow -\eta}$ is an eigenfunction of $H_2$, or
equivalently, that $\phi\arrowvert_{x_1\leftrightarrow x_2}$ is an
eigenfunction of $H_2\arrowvert_{\gamma\rightarrow
-\gamma,\eta\rightarrow -\eta}$. Note that $\phi$ and
$\phi\arrowvert_{x_1\leftrightarrow x_2, \gamma\rightarrow
-\gamma,\eta\rightarrow -\eta}$ are linearly independent. Since the
potential $\hat V_{12} = 0$ unless $x_1 = x_2$, every eigenfunction
$\psi$ of $H_2$ obeys the free Schr\"odinger equation
$\left(\partial_1^2 + \partial_2^2 + E\right)\psi(x_1,x_2) = 0$ in all
regions without coinciding coordinates. Thus, in these regions they
are linear combinations of the plane waves $\ee^{\ii k_1x_1 + \ii
k_2x_2}$ and $\ee^{\ii k_2x_1 + \ii k_1x_2}$. This implies that $\phi$
and $\phi\arrowvert_{x_1\leftrightarrow x_2, \gamma\rightarrow
-\gamma,\eta\rightarrow -\eta}$ provide, in the absence of bound
states, a complete set of eigenfunctions of $H_2$.

\subsection{Consistency conditions}
We now consider, for arbitrary $N$, the model defined by $H_N$ in
(\ref{HN}) and (\ref{Vjk}). In particular, we determine for which
values of the coupling constants $(c, \lambda, \gamma, \eta)$ its
eigenfunctions $\psi(x)$ can be obtained by the coordinate Bethe
Ansatz
\begin{equation}
\label{betheAnsatz}
	\psi(x) = \sum_{P\in
	S_N}A_P(Q)\ee^{\ii k_{P}\cdot x_Q}
\end{equation}
in the wedge
\begin{equation}
  \Delta_Q:\quad x_{Q(1)}<x_{Q(2)}<\ldots<x_{Q(N)}
\end{equation}
with $x = (x_1,\ldots,x_N)$ and $k_{P}\cdot x_Q =
\sum_{j=1}^Nk_{P(j)}x_{Q(j)}$, for all $Q\in S_N$ \cite{Y}. The
corresponding eigenvalue is obviously $E = \sum_{j=1}^Nk_j^2$.  We
recall that the validity of the Bethe Ansatz amounts to the model
being quantum integrable in the sense that the most that can happen in
any two-body scattering process is an exchange of particle momenta.

To take the boundary conditions (\ref{boundCond}) into account we
consider the boundary $x_j = x_k$ for fixed $j$ and $k$ such that
$j<k$. The last requirement is important since the particles in
general are non-identical, as previously discussed. Furthermore, let
$Q$ be an element in $S_N$ such that $x_{Q(i)}\equiv x_j$ and
$x_{Q(i+1)}\equiv x_k$ for some fixed $i$, implying that $x_j = x_k -
0^+$ is contained in the wedge $\Delta_Q$ and $x_j = x_k + 0^+$ in
$\Delta_{QT_i}$, where $T_i\in S_N$ is the transposition interchanging
$i$ and $i+1$. From this we deduce that the boundary conditions imply
the following relations between the coefficients $A_P(Q)$ of the
coordinate Bethe Ansatz (\ref{betheAnsatz}),
\begin{eqnarray}
  \ii\big(k_{P(i)} - k_{P(i+1)}\big)\big\lbrack A_{PT_i}(QT_i)
  - A_P(QT_i) - A_P(Q) + A_{PT_i}(Q)\big\rbrack = c\big\lbrack A_P(Q)
  \nonumber\\+\ A_{PT_i}(Q) + A_P(QT_i) + A_{PT_i}(QT_i)\big\rbrack -
  (\ii\gamma + \eta)\big\lbrack A_{PT_i}(QT_i) - A_P(QT_i)\nonumber\\+\
  A_P(Q) - A_{PT_i}(Q)\big\rbrack\nonumber\\ A_P(QT_i) +
  A_{PT_i}(QT_i) - A_P(Q) - A_{PT_i}(Q) =
  \ii\lambda\big(k_{P(i)} -
  k_{P(i+1)}\big)\big\lbrack A_{PT_i}(QT_i)\nonumber\\-\ A_P(QT_i) +
  A_P(Q) - A_{PT_i}(Q)\big\rbrack + (\gamma +
  \ii\eta)\big\lbrack A_P(Q) + A_{PT_i}(Q)\nonumber\\+\ A_P(QT_i) +
  A_{PT_i}(QT_i)\big\rbrack.
\end{eqnarray}
These relations constitute a linear, homogeneous system of $2(N -
1)N!^2$ equations for the $N!^2$ unknowns $A_P(Q)$, and the Bethe
Ansatz is consistent if and only if this over-determined system of
equations has $N!$ independent solutions where the $A_P(I)$ can be
chosen arbitrarily. In the following discussion we will show that this
is the case in the two special cases corresponding to the models
defined by (\ref{H1}) and (\ref{H2}). To do this we will not attempt
to solve these rather complicated system of equations by brute-force,
but rather use a somewhat indirect approach which essentially amounts
to reducing the $N$-particle case to a sequence of two particle
problems.

\begin{lemma}\label{ansatzLemma}
Let $Q\in S_N$ and $i\in \lbrace1,\ldots, N\rbrace$. If
$Q(i)<Q(i+1)$ and the coordinate Bethe Ansatz (\ref{betheAnsatz}) is
consistent, then
\begin{eqnarray}
\label{coeffRels}
	A_{PT_i}(Q) = S_R^+\left(k_{P(i)} - k_{P(i+1)}\right)A_P(Q) +
	S_T^-\left(k_{P(i)} - k_{P(i+1)}\right)A_P(QT_i)\nonumber\\
	A_{PT_i}(QT_i) = S_R^-\left(k_{P(i)} - k_{P(i+1)})A_P(QT_i) +
	S_T^+(k_{P(i)} - k_{P(i+1)}\right)A_P(Q),
\end{eqnarray}
where $T_i$ is the transposition interchanging $i$ and $i+1$.
\end{lemma}

\begin{proof}
We first consider the case $N = 2$. Let $\psi$ be
an arbitrary eigenfunction of $H_2$. Recall that the eigenfunctions
$\phi$ and $\phi\arrowvert_{x_1\leftrightarrow x_2, \gamma\rightarrow
-\gamma,\eta\rightarrow -\eta}$ of $H_2$ constitute, in the absence of
bound states, a complete basis for the eigenspace of $H_2$. It follows
that $\psi$ is a linear combination,
\begin{eqnarray}
	\psi & = & a_1\phi + a_2\phi\arrowvert_{x_1\leftrightarrow x_2,
	\gamma,\eta\rightarrow -\gamma,-\eta}\nonumber\\ & = &
	\left\{\begin{array}{ll} a_1\ee^{\ii k_1x_1 + \ii k_2x_2} + f_1(k_1 -
	k_2)\ee^{\ii k_2x_1 + \ii k_1x_2}, & x_1<x_2\\ f_2(k_1 -
	k_2)\ee^{\ii k_2x_2 + \ii k_1x_1} + a_2\ee^{\ii k_1x_2 + \ii k_2x_1}, &
	x_2<x_1 \end{array}\right.,
\end{eqnarray}
for some complex constants $a_1$ and $a_2$; we have introduced the
functions
\begin{eqnarray}
	f_1(k_1 - k_2) = S_R^+(k_1 - k_2)a_1 + S_T^-(k_1 -
	k_2)a_2\nonumber\\ f_2(k_1 - k_2) = S_R^-(k_1 -
	k_2)a_2 + S_T^+(k_1 - k_2)a_1.
\end{eqnarray}
Relabeling the constants $a_1$ and $a_2$ as well as the functions
$f_1$ and $f_2$ as follows, $a_1 = A_I(I)$, $a_2 = A_I(T_1)$,
$A_{T_1}(I) = f_1$ and $A_{T_1}(T_1) = f_2$, we arrive at the
two-particle Bethe Ansatz with coefficients $A_P(Q)$ given by
\begin{eqnarray}
\label{restriction}
	A_{T_1}(I) = S_R^+(k_1 - k_2)A_I(I) + S_T^-(k_1 -
	k_2)A_I(T_1)\nonumber\\ A_{T_1}(T_1) = S_R^-(k_1 - k_2)A_I(T_1)
	+ S_T^+(k_1 - k_2)A_I(I).
\end{eqnarray}

To extend this result to arbitrary values of $N$ we observe that the
boundary conditions (\ref{boundCond}) become identical with the
boundary conditions for the case $N = 2$ if we substitute
$x_j\rightarrow x_1$ and $x_k\rightarrow x_2$, for all $j < k$.  For
each $P\in S_N$, let $\Delta_Q$ be a wedge such that $x_{Qi}\equiv
x_j$ and $x_{Q(i+1)}\equiv x_k$. From the equivalence of boundary
conditions, in the sense stated above, follows that the relations
between the coefficients $A_P(Q)$, $A_P(QT_i)$, $A_{PT_i}(Q)$ and
$A_{PT_i}(QT_i)$ are obtained from (\ref{restriction}) by the
substitutions $x_1\rightarrow x_{Q(i)}$ and $x_2\rightarrow
x_{Q(i+1)}$ as well as $k_1\rightarrow k_{P(i)}$ and $k_2\rightarrow
k_{P(i+1)}$. This yields the relations in (\ref{coeffRels}).
\end{proof}

It is important to note that there is a possible inconsistency in the
coordinate Bethe Ansatz arising from the fact that the representation
of an element in $S_N$ is not unique. However, any two representations
can be converted into each other by using the defining relations of
$S_N$,
\begin{eqnarray}
\label{definingRels}
	T_iT_i = 1,\quad T_iT_{i + 1}T_i = T_{i + 1}T_iT_{i +
	1}\nonumber\\ T_iT_j = T_jT_i\quad \textrm{for}\ |i - j|>1.
\end{eqnarray}
Thus no inconsistency can arise provided that
\begin{eqnarray}
\label{consistencyConds}
	A_{PT_iT_i}(Q) = A_P(Q),\quad A_{PT_iT_{i+1}T_i}(Q) =
A_{PT_{i+1}T_iT_{i+1}}(Q)\nonumber\\ A_{PT_iT_j}(Q) =
A_{PT_jT_i}(Q)\quad \textrm{for} \ |i-j|>1
\end{eqnarray}
for all $P, Q\in S_N$.

To determine when these conditions are valid we follow the approach of
A. B. and Al. B. Zamolodchikov \cite{Z} and make use
of the algebraic structure which has come to be known as the {\em
Zamolodchikov algebra}. We start by briefly recalling the construction of the
Zamolodchikov algebra: to each particle of type $A$ and with momenta
$k$ the symbol $A(k)$ is associated, and the two particle scattering
theory is encoded in the commutation relation
\begin{equation}
  A(k_1)B(k_2) = S_R^{AB}(k_{12})A(k_2)B(k_1) +
  S_T^{AB}(k_{12})B(k_2)A(k_1),
\end{equation}
where $k_{12} = k_1 - k_2$ and $S_R^{AB}$ and $S_T^{AB}$ are the two
particle scattering amplitudes. The full $N$ particle scattering
theory is then obtained by factorizing each scattering event into a
product of two particle events. As observed in \cite{Z}, identifying
each product $A(k_1)B(k_2)C(k_3)\ldots$ with a particular coefficient
$A_P(Q)$ of the coordinate Bethe Ansatz the consistency conditions
(\ref{consistencyConds}) are equivalent to requiring the Zamolodchikov
algebra to be consistent as well as associative. In our particular
case, since the particles are distinguished only by their relative
ordering, it is sufficient to consider the case of three
particles. Therefore, let the product $A(k_1)B(k_2)C(k_3)$ correspond
to a particular coefficient $A_P(Q)$, for fixed $P,Q\in S_3$. A
straightforward but somewhat tedious computation then shows that the
Zamolodchikov algebra, in our case, is consistent as well as
associative if and only if the two-particle scattering amplitudes
$S_R^{\pm}$ and $S_T^{\pm}$ obey the following so called {\em
Factorization equations},
\begin{eqnarray}
\label{factor1}
        S_R^+(u)S_R^+(-u) + S_T^-(u)S_T^+(-u) = 1\\
	S_R^-(u)S_R^-(-u) + S_T^+(u)S_T^-(-u) = 1\\
	S_R^+(u)S_T^-(-u) + S_T^-(u)S_R^-(-u) = 0\\
\label{factor2}
	S_R^-(u)S_T^+(-u) + S_T^+(u)S_R^+(-u) = 0\\
\label{factor3}
        S_R^-(v)S_R^+(u + v)S_R^-(u) = S_R^+(u)S_R^-(u + v)S_R^+(v)\\
\label{factor4}
	S_R^+(v)S_T^+(u + v)S_T^-(u) = S_T^+(u)S_T^-(u + v)S_R^+(v)\\
\label{factor5}
	S_R^-(v)S_T^-(u + v)S_T^+(u) = S_T^-(u)S_T^+(u + v)S_R^-(v)\\
\label{factor6}
	S_R^+(v)S_R^+(u + v)S_T^-(u) + S_T^-(v)S_R^+(u + v)S_R^-(u) =
	S_R^+(u)S_T^-(u + v)S_R^+(v)\\ S_R^-(v)S_R^+(u + v)S_T^+(u) +
	S_T^+(v)S_R^+(u + v)S_R^+(u) = S_R^+(u)S_T^+(u + v)S_R^+(v)\\
	S_R^-(v)S_R^-(u + v)S_T^+(u) + S_T^+(v)S_R^-(u + v)S_R^+(u) =
	S_R^-(u)S_T^+(u + v)S_R^-(v)\\ S_R^+(v)S_R^-(u + v)S_T^-(u) +
	S_T^-(v)S_R^+(u + v)S_R^-(u) = S_R^-(u)S_T^-(u + v)S_R^+(v)\\
	S_R^+(v)S_R^-(u + v)S_T^-(u) + S_T^-(v)S_R^-(u + v)S_R^-(u) =
	S_R^-(u)S_T^-(u + v)S_R^-(v)\\
\label{factor7}
	S_R^-(v)S_R^+(u + v)S_T^+(u) + S_T^+(v)S_R^-(u + v)S_R^+(u) =
	S_R^+(u)S_T^+(u + v)S_R^-(v)
\end{eqnarray}
for all real $u$ and $v$.

Upon substituting the two-particle scattering amplitudes
(\ref{scattAmpl}) and (\ref{scattAmplMinus}) into the Factorization
equations above a straightforward but somewhat tedious calculation
shows that (\ref{factor1})--(\ref{factor2}) as well as
(\ref{factor4})--(\ref{factor5}) are fulfilled for all values of $(c,
\lambda, \gamma, \eta)$, while (\ref{factor3}) and
(\ref{factor6})--(\ref{factor7}) holds true if and only if
\begin{eqnarray}
  \gamma\lbrack\lambda(3c + \lambda(u^2 + uv + v^2)) +
  4\gamma^2\rbrack = 0\nonumber\\ \lbrack -1 + c\lambda + \gamma^2 +
  \eta(\eta - 2\ii)\rbrack \lbrack\lambda(3c + \lambda(u^2 + uv + v^2))
  + 4\gamma^2\rbrack = 0\nonumber\\ \lbrack -1 + c\lambda + \gamma^2 +
  \eta(\eta - 2\ii)\rbrack \lbrack\lambda(3c + \lambda(u^2 + uv + v^2))
  + 4\gamma^2\rbrack = 0
\end{eqnarray}
for all real $u$ and $v$. Using the fact that the first condition requires
$\gamma = 0$ and that the last two are related by complex conjugation, 
we can reduce them to the rather simple form
\begin{eqnarray}
  \gamma = 0\nonumber\\ \lambda\lbrack -1 + c\lambda + \eta(\eta -
  2\ii)\rbrack = 0,
\end{eqnarray}
which obviously has two families of solutions, each of which
defines systems exactly solvable by the coordinate Bethe
Ansatz. We thus arrive at the main result of this paper.

\begin{thm}\label{mainThm}
The coordinate Bethe Ansatz (\ref{betheAnsatz}) for the eigenfunctions
of the Hamiltonian $H_N$ is consistent if and only if
\begin{equation}
\label{thmCondOne}
  \lambda = \gamma = 0
\end{equation}
or
\begin{equation}
\label{thmCondTwo}
  \lambda = 1/c\quad\textrm{and}\quad \gamma = \eta = 0 . 
\end{equation}
\end{thm}

Note that these two sets of conditions on the coupling constants
$(c,\lambda,\gamma,\eta)$ correspond to the two Hamiltonians defined
in (\ref{H1}) and (\ref{H2}). It is interesting to note that, in both
cases, $S_R^\pm$ and $S_T^\pm$ only have a single pole, and
$S_R^+=S_R^-$ even in the first case.

\subsection{Explicit results for the case $(c, 1/c, 0, 0)$}
This particular case correspond to the Hamiltonian $H^{(2)}$ in
Eq. (\ref{H2}). We note that this Hamiltonian is invariant under
permutations of the coordinates $x_j$, and one can therefore assume
a particular exchange statistics, which determines the eigenfunction
in all wedges once it is known in one of them.  Furthermore, using the
fact that $S_T^+\arrowvert_{\lambda=1/c,\gamma=\eta=0} =
S_T^-\arrowvert_{\lambda=1/c,\gamma=\eta=0} = 0$ and
$S_R^+\arrowvert_{\lambda=1/c,\gamma=\eta=0} =
S_R^-\arrowvert_{\lambda=1/c,\gamma=\eta=0}$, the relations in
(\ref{coeffRels}) reduce to the rather simple form
\begin{eqnarray}
	A_{PT_i}(Q) & = &
	S_R^+\arrowvert_{\lambda=1/c,\gamma=\eta=0}\left(k_{P(i)} -
	k_{P(i + 1)}\right)A_P(Q)\nonumber\\ & = &
	\frac{\ii\left(k_{P(i)} - k_{P(i+1)}\right)^2 +
	\ii c^2}{\ii\left(k_{P(i)} - k_{P(i+1)}\right)^2 + 2\left(k_{P(i)}
	- k_{P(i+1)}\right) - \ii c^2}A_P(Q)\nonumber\\ & = &
	\frac{\ii\left(k_{P(i)} - k_{P(i+1)}\right) - c}{\ii\left(k_{P(i)}
	- k_{P(i+1)}\right) + c}A_P(Q)
\end{eqnarray}
for all $Q\in S_N$. This implies that
\begin{equation}
	A_{PT_i} = Y_i\left(k_{P(i)} - k_{P(i+1)}\right)A_P,
\end{equation}
where we have introduced the function
\begin{equation}
	Y_i(u) = \frac{\ii u + c}{\ii u - c}, 
\end{equation}
and interpret $A_P$ as a vector with $N!$ elements $A_P(Q)$. It is
important to note that the functions $Y_i$, for arbitrary exchange
statistics, are identical to the corresponding functions appearing
when restricting the delta-interaction model to bosons (see
e.g. \cite{chang1995}).  The eigenfunctions of the latter model are
well-known (see e.g. Section I.1 in \cite{korepin1993})
\begin{equation}
	\psi = C\left\lbrack \prod_{N\geq j>k\geq 1}(\partial_{x_j} -
	\partial_{x_k} + c)\right\rbrack \det_{1\leq m,n\leq N}\lbrack
	\textrm{exp}(\ii k_mx_n)\rbrack,\quad x_1<x_2<\ldots<x_N
\end{equation}
where $C$ is a normalization constant. We conclude that this formula
also gives the eigenfunctions of the Hamiltonian in (\ref{H2}) in the
identity wedge $\Delta_I$, for arbitrary exchange statistics
(characterized by some Young tableaux). Using standard arguments from
group theory (see e.g. \cite{W}), these eigenfunctions can be
straightforwardly extended to all other wedges.

\subsection{Recursion relations for the coefficients $A_P$}
In order to provide a machinery for computing the eigenfunctions
explicitly we now proceed to derive a set of recursion relations for
the coefficients $A_P$ of the coordinate Bethe Ansatz. Our starting
point is the following fact: defining
\begin{equation}
\label{opDef}
  (\hat R)_{Q, Q^{\prime}} = \delta_{Q^{\prime},QR}
\end{equation}
one can write
\begin{equation}
  A_P(QR) = \sum_{Q^{\prime}\in W_N}(\hat R)_{Q,
  Q^{\prime}}A_P(Q^{\prime}) = (\hat RA_P)(Q),
\end{equation}
where the first equality is a trivial consequence of the definition,
and in the second we interpret $(\hat R)_{Q, Q^{\prime}}$ as elements
of an $n\times n$ matrix $\hat R$ with $n = N!$ the rank $|S_N|$ of
$S_N$.  These matrices obviously define a representation $R\to \hat R$
of $S_N$ acting on the coefficients $A_P(Q)$. It is worth noting that
this is identical with the so called (right) regular representation of
$S_N$. Using this fact we can rewrite the two relations in
(\ref{coeffRels}) as follows,
\begin{eqnarray}
\label{recursRels}
  A_{PT_i}(Q) = \left\lbrack S_R^+\left(k_{P(i)} - k_{P(i+1)}\right) +
  S_T^-\left(k_{P(i)} - k_{P(i+1)}\right)\hat T_i\right\rbrack
  A_P(Q)\nonumber\\ A_{PT_i}(QT_i) = \left\lbrack S_R^-\left(k_{P(i)}
  - k_{P(i+1)}\right) + S_T^+\left(k_{P(i)} - k_{P(i+1)}\right)\hat
  T_i\right\rbrack A_P(QT_i),
\end{eqnarray}
where $Q$ is required to fulfill the condition $Q(i) < Q(i + 1)$. To
organize the elements of the vector $A_P$ we have to introduce an
ordering on the set of permutations. We find it convenient to use the
following

\begin{definition}\label{ordDef}
Associate to any two permutations $Q,Q^{\prime}\in S_N$ the sequence
$a_i := Q(N - i + 1) - Q^{\prime}(N - i + 1)$, where
$i=1,2,\ldots,N$. If the first non-zero number in the sequence
$\lbrace a_i\rbrace$ is positive, $Q$ is said to be larger than
$Q^{\prime}$, denoted as $Q>Q^{\prime}$.
\end{definition}

The idea for the remainder is the following: order the coefficients
$A_P(Q)$ according to the ordering just defined (largest permutation
first) into the vector $A_P$, and use the recursion relations in
(\ref{recursRels}) for the individual coefficients $A_P(Q)$ to write
\begin{equation}
\label{recursRel}
  A_{PT_i} = \mathbf{Y}_i\left(k_{P(i)} - k_{P(i+1)}\right)A_P,
\end{equation}
where we have introduced the matrix
\begin{equation}
\label{matrixYOp}
  \mathbf{Y}_i(u) = \mathbf{S}^i_R(u) + \mathbf{S}^i_T(u)\hat T_i
\end{equation}
in which $\mathbf{S}^i_R$ and $\mathbf{S}^i_T$ are diagonal $N!\times
N!$ matrices with entries $S_R^{\pm}$ and $S_T^{\pm}$
respectively. To determine the distribution of the scattering
amplitudes $S_R^{\pm}$ and $S_T^{\pm}$ among the diagonal elements of
the matrices $\mathbf{S}_R^i$ and $\mathbf{S}_T^i$ we start by
deducing a natural decomposition of an arbitrary permutation into a
product of elementary transpositions $T_i$. This decomposition will
provide us with enough information to determine the structure of
$\mathbf{S}_R^i$ and $\mathbf{S}_T^i$ explicitly.

We pause to introduce some notation and conventions to be used in the
remainder of the discussion. By abuse of notation, we identify a
permutation $Q\in S_N$ with $Q^{\prime}\in S_{N+n}$, $n \geq 0$, if
\begin{eqnarray}
  Q(i) = Q^{\prime}(i)\quad \textrm{for all}\ i = 1, 2, \ldots,
  N\nonumber\\ Q^{\prime}(i) = i\quad \textrm{for all}\ i = N+1,
  N+2,\ldots, n.
\end{eqnarray}
E.g., $(231)\in S_3$ and $(23145)\in S_5$ will be identified.\footnote{We
recall that a permutation $Q=(ijk)$ is defined such that $Q(1)=i$,
$Q(2)=j$ and $Q(3)=k$, etc.}  Furthermore, we denote as $Q^m_k$ the
$k$th permutation of $S_m$, ordered according to the ordering defined
in Definition \ref{ordDef}. E.g. for $S_3$ this implies that $Q^3_1 =
(123)$, $Q^3_2 = (213)$, $Q^3_3 = (132)$, $Q^3_4 = (312)$, $Q^3_5 =
(231)$ and $Q^3_6 = (321)$.

Using the fact that every integer $j$ such that $1\leq j\leq m!$ can
be uniquely written in the form $j = n(m-1)! + k$, where $0\leq n\leq
m-1$ and $1\leq k\leq (m-1)!$, we prove

\begin{lemma}\label{decompLemma}
Let $k$, $m$ and $n$ be integers such that $m>1$, $0\leq n\leq m-1$
and $1\leq k\leq (m-1)!$. Then
\begin{equation}
\label{permRecurs}
  Q^m_{n(m-1)!+k} =
  \prod_{i=m-n}^{\stackrel{m-1}{\longrightarrow}}T_i\ Q_{k}^{m-1},
\end{equation}
where
\begin{equation}
  \prod_{i=j}^{\stackrel{k}{\longrightarrow}}T_i =
  \left\lbrace\begin{array}{ll}T_jT_{j+1}\ldots T_k, & j\leq k\\ 0 &
  \textrm{otherwise}\end{array}\right..
\end{equation}
\end{lemma}

\begin{proof}
It clearly follows from Definition \ref{ordDef} and the fact
$|S_m| = m!$ that $Q^m_{n(m-1)!+1}$ is obtained by cyclically
permuting the last $m-n$ elements of the identity permutation in $S_m$
such that $Q^m_{n(m-1)!+1}(m) = m-n$. Decomposing this cyclic
permutation into a product of elementary transpositions $T_i$, we
obtain
\begin{equation}
  Q^m_{n(m-1)!+1} = T_{m-n}\ldots T_{m-1}.
\end{equation}
We conclude the proof of the Lemma by observing that
\begin{equation}
  Q^m_{n(m-1)!+k} = Q^m_{n(m-1)!+1}Q^{m-1}_{k}.
\end{equation}
\end{proof}

\begin{cor}\label{corDecompLemma}
Let $k$, $m$ and $n$ be integers such that $m>1$, $0\leq n\leq m$
and $1\leq k\leq m!$. Then
\begin{eqnarray}
  Q^{m+1}_{nm!+k}(m) > Q^{m+1}_{nm!+k}(m+1),\quad 1\leq k\leq n(m -
  1)!\nonumber\\ Q^{m+1}_{nm!+k}(m) < Q^{m+1}_{nm!+k}(m+1),\quad
  n(m - 1)!< k\leq m!.
\end{eqnarray}
\end{cor}

\begin{proof}
{}From Lemma \ref{decompLemma} clearly follows that
$Q^{m+1}_{nm!+1}(m+1) = m+1-n$, and consequently that
\begin{eqnarray}
  Q^{m+1}_{nm!+1}(j)<Q^{m+1}_{nm!+1}(m+1),\quad 1\leq j\leq
  m-n\nonumber\\ Q^{m+1}_{nm!+1}(j)>Q^{m+1}_{nm!+1}(m+1),\quad m-n<
  j\leq m.
\end{eqnarray}
In other words, for each fixed $n$ there exists $n$ distinct integers
$l\leq m+1$ such that $l>Q^{m+1}_{nm!+1}(m+1)$. This, together with
the facts that $Q^{m+1}_{nm!+k} = Q^{m+1}_{nm!+1}Q^{m}_{k}$ and that
$Q^{m}_{k}(m) = m$, for all $k\leq (m-1)!$, imply the statement.
\end{proof}

Note that Lemma \ref{decompLemma} provides a recursive procedure for
decomposing an arbitrary permutation into a product of elementary
transpositions $T_i$.

Using Lemma \ref{decompLemma} and Corollary \ref{corDecompLemma} we
now determine the structure of the matrices $\mathbf{S}_R^i$ and
$\mathbf{S}_T^i$. Corollary \ref{corDecompLemma} and the recursion
relations (\ref{recursRels}) together imply that
\begin{equation}
\label{SR}
  \left(\mathbf{S}^i_R\right)_{jj} = \left\lbrace\begin{array}{ll}
  S^-_R & 1\leq j - ni!\leq n(i - 1)!\\ S^+_R & n(i - 1)!< j - ni!\leq
  i!
  \end{array}\right.
\end{equation}
and
\begin{equation}
\label{ST}
  \left(\mathbf{S}^i_T\right)_{jj} = \left\lbrace\begin{array}{ll}
  S^+_T & 1\leq j - ni!\leq n(i - 1)!\\ S^-_T & n(i - 1)!< j - ni!\leq
  i!
  \end{array}\right.,
\end{equation}
where $n$ is required to fulfill the relation $j = ni! + k$, for some
$1\leq k\leq i!$. This determines only the first $(i + 1)!$ diagonal
elements of the $N!\times N!$ matrices $\mathbf{S}^i_R$ and
$\mathbf{S}^i_T$. However, it immediately follows from Lemma
\ref{decompLemma} that this structure is periodic with period $(i +
1)!$, i.e.,
\begin{eqnarray}
\label{SPeriod}
  \left(\mathbf{S}^i_R\right)_{j+(i + 1)!,j+(i + 1)!} =
  \left(\mathbf{S}^i_R\right)_{jj}\nonumber\\
  \left(\mathbf{S}^i_T\right)_{j+(i + 1)!,j+(i + 1)!} =
  \left(\mathbf{S}^i_T\right)_{jj},
\end{eqnarray}
thus determining all $N!^2$ elements of $\mathbf{S}^i_R$ and
$\mathbf{S}^i_T$. We recall that for our case,
\begin{eqnarray}
\label{STSR}
	S_T^\pm(u) = \frac{(-\eta^2 \pm 2 \ii \eta  + 
	1)u}{ (\eta^2 + 1)u + 
	\ii c},\quad  S_R^\pm(u) = \frac{-
	\ii c}{(\eta^2 + 1)u + \ii c},
\end{eqnarray}
but we give the construction for the more general case where $S^\pm_R$
are different since we hope that this will be useful for other models.

In deriving the recursion relations (\ref{recursRel}) it is important
to note that there is a possible inconsistency in the coordinate Bethe
Ansatz, arising from the fact that the elementary transpositions $T_i$
obey the defining relations (\ref{definingRels}) of the permutation
group $S_N$, as discussed in Section 3.2. Thus the recursion relations
(\ref{recursRel}) are consistent if and only if
\begin{eqnarray}
\label{vectorConsCond}
  A_{PT_iT_i} = A_P,\quad A_{PT_iT_{i+1}T_i} =
  A_{PT_{i+1}T_iT_{i+1}},\nonumber\\ A_{PT_iT_j} =
  A_{PT_jT_i}\quad \textrm{for} \ |i-j|>1
\end{eqnarray}
for all $P\in S_N$. Using the recursion relations (\ref{recursRel})
one find that that these latter conditions holds true if and only if
the following so called Yang-Baxter relations are fulfilled,
\begin{eqnarray}
\label{mtrxRels}
	\mathbf{Y}_i(-u)\mathbf{Y}_i(u) = I,\quad
	\mathbf{Y}_i(v)\mathbf{Y}_{i+1}(u + v)\mathbf{Y}_i(u) =
	\mathbf{Y}_{i+1}(u)\mathbf{Y}_i(u + v)\mathbf{Y}_{i+1}(v),
	\nonumber\\ \mathbf{Y}_i(u)\mathbf{Y}_j(v) =
	\mathbf{Y}_j(v)\mathbf{Y}_i(u)\quad \textrm{for} \ |i - j|>1
\end{eqnarray}
for all real $u$ and $v$. However, since we already in Section 3.2 determined
for which values of the coupling parameters $(c,\lambda,\gamma,\eta)$
the relations (\ref{vectorConsCond}) holds true we can, in view of Theorem
\ref{mainThm}, immediately conclude that we have proven
\begin{prop}
If and only if the coupling parameters $(c,\lambda,\gamma,\eta)$
satisfy the conditions in (\ref{thmCondOne}) or (\ref{thmCondTwo}),
then the Yang-Baxter relations (\ref{mtrxRels}) are fulfilled, and
\begin{equation}
  A_P = \mathbf{\mathcal{Y}}_P(k)A_I,
\end{equation}
where $\mathbf{\mathcal{Y}}_P(k)$ is a product of the matrices
$\mathbf{Y}_i(k_{P(i)} - k_{P(i+1)})$ obtained by repeatedly using
(\ref{recursRel}).
\end{prop}

\begin{rmk}
In the special case $\gamma = \eta = 0$ the Hamiltonian $H_N$ is
invariant under permutations of the coordinates, and in consequence
the scattering amplitude $S_R^+\arrowvert_{\gamma=\eta=0} =
S_R^-\arrowvert_{\gamma=\eta=0} =: S_R$ and similarly
$S_T^+\arrowvert_{\gamma=\eta=0} = S_T^-\arrowvert_{\gamma=\eta=0} =:
S_T$. A particularly interesting special case is the delta interaction
model $\lambda = \gamma = \eta = 0$, in which
case the matrices $\mathbf{Y}_i$ take the well known form
\begin{equation}
  \mathbf{Y}_i(u)\arrowvert_{\lambda=\gamma=\eta=0} = \frac{\ii u\hat
  T_i + c\hat I}{\ii u - c},
\end{equation}
originally introduced by Yang \cite{Y}. 
\end{rmk}

To illustrate the general discussion of this section we provide in
Appendix A a more explicit account of the three particle case $N=3$.
In Appendix B we outline a simple, direct proof of the Yang-Baxter
equations for the cases stated in Proposition 3.6.

\section{Concluding remarks}
In the introduction we argued that the Hamiltonian $H_N$ defined in
(\ref{HN}) and (\ref{Vjk}) is the $N$ particle generalization of the
most general Hamiltonian $H = -\partial_x^2 + \hat V$ with local
interaction $\hat V$. However, not all $N$-body Hamiltonians with
local {\em two-body} interactions can be obtained in this way. A
particular example of such a model is the quantum version of the
derivative nonlinear Schr\"odinger equation \cite{NS} defined by the
Hamiltonian
\begin{equation}
  H_N = -\sum_{j=1}^N\partial_{x_j}^2 +
  2 \tilde \eta\sum_{j<k}\delta(x_j - x_k)\ii \left(\partial_{x_j} +
  \partial_{x_k}\right),
\end{equation}
where $\tilde\eta$ is a real coupling parameter. This model does not
fall in the class studied in this paper since the interaction depends
also on the sum of the momenta of the particles involved in the
interaction, and it is therefore not Galilean invariant.  However, this
Hamiltonian is nevertheless interesting. It describes identical
particles, and it was shown by Gutkin \cite{G} that it is exactly
solvable by the coordinate Bethe Ansatz only in the case of bosons or
fermions.

All possible local two-body interactions are formally defined by the
2-body Hamiltonian $H_2=-\partial_{x_1}^2-\partial_{x_2}^2+\hat W_{12}$,
where $(\hat W_{12}\psi)(x_1,x_2)=0$ for two-particle wave functions
$\psi$ except in regions of coinciding coordinates, $x_1=x_2$. We
believe that there is a nine parameter family of such two-body
interactions which one should be able to find using general methods
discussed in \cite{AGHH,AK}. It would be interesting to know if there
are additional such distinguishable particle models exactly solvable
by the coordinate Bethe Ansatz.

As discussed in \cite{EG} Remark~2.4, there is a one parameter family
of local interactions which are unitarily equivalent to the
non-interacting case. This suggests that there should be also a one parameter
extension of local interactions which are unitarily
equivalent to the delta-interaction, and this might provide a simple
explanation of our results. At closer inspection we found that this is
indeed the case: if and only if $\psi(x)$ obeys the boundary
conditions in (\ref{bcgeneral}) defining the interaction
$(c,0,\eta,0)$, then
\begin{eqnarray} 
\tilde\psi(x) = \ee^{-\ii\alpha\Theta(x)}\psi(x),\quad \ee^{\ii\alpha}
\equiv \frac{1+\ii\eta}{1-\ii\eta}
\end{eqnarray} 
with the Heaviside function $\Theta$ obeys the boundary conditions
defining the interaction $(c/(1+\eta^2),0,0,0)$. In a similar manner,
the model formally defined in (\ref{H1}) is unitarily equivalent to
the delta-gas with coupling $\tilde c = c/(1+\eta^2)$, and the unitary
operator $U$ intertwining the two models is given by
\begin{eqnarray} 
(U\psi)(x_1,\ldots,x_N) = \ee^{-\ii\alpha\sum_{j<k}\Theta(x_j-x_k)}
\psi(x_1,\ldots,x_N)
\end{eqnarray} 
with $\alpha$ as above.\footnote{We stress that this unitary
equivalence cannot be understood on the level of formal Hamiltonians
since our local interactions include non-trivial regularizations.}
Moreover, it is interesting to note that (\ref{STSR}) can be written
as
\begin{eqnarray} 
	S_T^\pm(u) = \ee^{\pm \ii \alpha}b(u), \quad S_R^\pm (u) =a(u) 
\end{eqnarray} 
where $Y_i(u) = a(u) + b(u)T_i$ gives the rational solution of the
Yang-Baxter relations, suggesting that our solution $\mathbf{Y}_i(u)$
of the Yang-Baxter relations is unitary equivalent to the latter.  We
should mention that this very argument was used already in \cite{AFK1}
to deduce the integrability of the model in (\ref{H1}).\footnote{We
thank P. Kurasov for pointing this out to us.} However, while this is
a simple alternative argument showing integrability of the model in
(\ref{H1}), only our argument allows to conclusively determine {\em
all} integrable cases.

To conclude, in this paper we found an example of a 1D quantum
many-body system of {\em non-identical} particles which is integrable,
and it led us to an interesting class of solutions of the Yang-Baxter
relations. While the solution which we found is closely related to the
well-known rational one, there might be others which are truly
different from known ones, and our results provide a convenient
starting point for finding them. It thus is tempting to speculate that
our model is only a first example in a novel interesting class of
integrable systems.

\section*{Acknowledgements}
We would like to thank Harald Grosse for asking us the question
answered in this paper, as well as for his encouragement along the
many phases of this project. We would also like to thank P. Kulish for
several helpful remarks. 
C.P. was supported by {\em Deutsche
Forschungsgemeinschaft} under the Emmy-Noether Programme. E.L. was
supported in part by the Swedish Science Research Council~(VR) and the
G\"oran Gustafsson Foundation.

\app
\section*{Appendix A. Recursion relations for the three particle case}
To illustrate the abstract discussion of Section 3.4 we here provide
explicit formulas for the three particle case. We start by noting that
the vector $A_P$, in the ordering defined in Section 3.3, has the
following structure,
\begin{equation}
  A_P^t = {\big (}A_P(123),\ A_P(213),\ A_P(132),\ A_P(312),\
  A_P(231),\ A_P(321){\big )},
\end{equation}
where $A_P^t$ denotes the transpose of $A_P$. Straightforward
computations (or alternatively Lemma \ref{decompLemma}) show that we
can write this using the elementary transpositions $T_1$ and $T_2$
as follows,
\begin{equation}
  A_P^t = {\big (}A_P(I),\ A_P(T_1),\ A_P(T_2),\ A_P(T_1T_2),\
  A_P(T_2T_1),\ A_P(T_1T_2T_1){\big )}.
\end{equation}
Directly applying the recursion relations (\ref{recursRels}) we
find that the matrices $S_R^1$ and $S_T^1$, in the particular ordering
we have chosen, take the following form,
\begin{equation}
  \mathbf{S}_R^1 = \left(\begin{matrix}
    S_R^+ & 0 & 0 & 0 & 0 & 0\\
    0 & S_R^- & 0 & 0 & 0 & 0\\
    0 & 0 & S_R^+ & 0 & 0 & 0\\
    0 & 0 & 0 & S_R^- & 0 & 0\\
    0 & 0 & 0 & 0 & S_R^+ & 0\\
    0 & 0 & 0 & 0 & 0 & S_R^-
  \end{matrix}\right),\quad 
  \mathbf{S}_T^1 = \left(\begin{matrix}
    S_T^- & 0 & 0 & 0 & 0 & 0\\
    0 & S_T^+ & 0 & 0 & 0 & 0\\
    0 & 0 & S_T^- & 0 & 0 & 0\\
    0 & 0 & 0 & S_T^+ & 0 & 0\\
    0 & 0 & 0 & 0 & S_T^- & 0\\
    0 & 0 & 0 & 0 & 0 & S_T^+
  \end{matrix}\right).
\end{equation}
Using $T_1T_2T_1 = T_2T_1T_2$ and (\ref{recursRels}) we deduce that
\begin{equation}
  \mathbf{S}_R^2 = \left(\begin{matrix}
    S_R^+ & 0 & 0 & 0 & 0 & 0\\
    0 & S_R^+ & 0 & 0 & 0 & 0\\
    0 & 0 & S_R^- & 0 & 0 & 0\\
    0 & 0 & 0 & S_R^+ & 0 & 0\\
    0 & 0 & 0 & 0 & S_R^- & 0\\
    0 & 0 & 0 & 0 & 0 & S_R^-
  \end{matrix}\right),\quad 
  \mathbf{S}_T^2 = \left(\begin{matrix}
    S_T^- & 0 & 0 & 0 & 0 & 0\\
    0 & S_T^- & 0 & 0 & 0 & 0\\
    0 & 0 & S_T^+ & 0 & 0 & 0\\
    0 & 0 & 0 & S_T^- & 0 & 0\\
    0 & 0 & 0 & 0 & S_T^+ & 0\\
    0 & 0 & 0 & 0 & 0 & S_T^+
  \end{matrix}\right).
\end{equation}
It is straightforward to verify that this structure is indeed
reproduced also by the general discussion presented in Section 3.3, in
particular Eqs. (\ref{SR})--(\ref{SPeriod}). To explicitly construct
the matrices $\mathbf{Y}_i$, for $i = 1, 2$, and thereby also the
recursion relations (\ref{recursRel}), there remains only to determine
the explicit form of $\hat T_1$ and $\hat T_2$. Using (\ref{opDef}) we
find that
\begin{equation}
  \hat T_1 = \left(\begin{matrix}
    0 & 1 & 0 & 0 & 0 & 0\\
    1 & 0 & 0 & 0 & 0 & 0\\
    0 & 0 & 0 & 1 & 0 & 0\\
    0 & 0 & 1 & 0 & 0 & 0\\
    0 & 0 & 0 & 0 & 0 & 1\\
    0 & 0 & 0 & 0 & 1 & 0
  \end{matrix}\right),\quad 
  \hat T_2 = \left(\begin{matrix}
    0 & 0 & 1 & 0 & 0 & 0\\
    0 & 0 & 0 & 0 & 1 & 0\\
    1 & 0 & 0 & 0 & 0 & 0\\
    0 & 0 & 0 & 0 & 0 & 1\\
    0 & 1 & 0 & 0 & 0 & 0\\
    0 & 0 & 0 & 1 & 0 & 0
  \end{matrix}\right).
\end{equation}
Inserting the above matrices into the consistency conditions
(\ref{mtrxRels}) a straightforward but somewhat tedious calculation
shows that they are equivalent to Factorization equations
(\ref{factor1})--(\ref{factor7}). This reflects the fact that, for
this model, it is sufficient to consider the three-particle case in
order to find those cases exactly solvable by the coordinate Bethe
Ansatz.
\appende

\app
\section*{Appendix B. Direct proof of Yang-Baxter equations}
In this Appendix we outline an alternative, direct proof of the
validity of the Yang-Baxter relations in (\ref{mtrxRels}) for the
operators $\mathbf{Y}_i(u)$ defined in Eqs.\ (\ref{matrixYOp}) and
(\ref{scattAmpl})--(\ref{scattAmplMinus}) and the cases stated in
Theorem~3.2.

We first note that the formulas given in Appendix~A allow a simple
verification of these relations in the simplest non-trivial case
$N=3$.\footnote{The verification can easily be performed by using a symbolic
programming language like MAPLE or MATHEMATICA.} The general result, for
arbitrary $N$, then
follows from the following

\begin{lemma} If the operators $\mathbf{Y}_i(u)$ satisfy the Yang-Baxter
relations 
for $N=3$ they satisfy them for any $N>2$.
\end{lemma}

This is true since all identities in (\ref{mtrxRels}) can be brought to a block
diagonal form, where the relations in each block are unitarily
equivalent to the corresponding ones for the case $N = 3$.

To be more specific, we consider first the second identity in
(\ref{mtrxRels}), involving three matrices $\mathbf{Y}$ on each side,
for some fixed $N>2$ and $i<N$. It follows from (\ref{matrixYOp}) that
$\mathbf{Y}_i(u)$ acting on $A_P$ mixes only elements $A_P(Q)$ and
$A_P(Q T_i)$ (since $\mathbf{S}^i_{R,T}$ are diagonal matrices). Thus,
for any $Q\in S_N$, both sides of the second set of identities in
(\ref{mtrxRels}) mix only the elements $A_P(Q)$, $A_P(QT_i)$,
$A_P(QT_{i+1})$, $A_P(QT_{i+1}T_i)$ and $A_P(QT_i T_{i+1})$, i.e., it
is possible to reorder the elements of the vector $A_P$ so that one
can group them into blocks of six elements such that the identities
decompose into blocks of $6\times 6$ matrices. Moreover, let $Q'$ be
the largest element, when using the ordering defined in Definition
\ref{ordDef}, in the set
$\{Q,QT_i,QT_{i+1},QT_{i+1}T_i,QT_iT_{i+1},Q_iT_{i+1}T_i\}$, then
\begin{equation}
  Q'>Q'T_i>Q'T_{i+1}>Q'T_{i+1}T_i>Q'T_iT_{i+1}>Q'T_iT_{i+1}T_i . 
\end{equation}
This implies that the matrices $\mathbf{Y}_i(u)$ and
$\mathbf{Y}_{i+1}(u)$ restricted to the vector
\begin{equation} 
  \bigl( A_P(Q'),A_P(Q'T_i), A_P(Q'T_{i+1}), A_P(Q'T_{i+1}T_i),
  A_P(Q'T_iT_{i+1}), A_P(Q'T_iT_{i+1}T_i) \bigr)^t 
\end{equation}
are identical with the $6\times 6$ matrices $\mathbf{Y}_1(u)$ and
$\mathbf{Y}_{2}(u)$ for $N=3$ respectively (these latter matrices are
explicitly given in Appendix~A). This proves the lemma for the second
set of identities in (\ref{mtrxRels}).  The verification of the first
and last set of identities follows from similar arguments (they reduce to
blocks of $2\times 2$ and $4\times 4$ matrix identities which are easy
to verify).
\appende

\end{document}